\documentclass[11pt,preprintnumbers,aps,amssymb,nofootinbib,amsmath]
{revtex4}
\usepackage{epsfig,epsf}
\usepackage{bm} 
\newcommand{\beq}{\begin{equation}}
\newcommand{\beql}[1]{\begin{equation}\label{#1}}
\newcommand{\eeq}{\end{equation}}
\newcommand{\bea}{\begin{eqnarray}}
\newcommand{\eea}{\end{eqnarray}}
\newcommand{\eq}[1]{(\ref{#1})}
\newcommand{\fig}[1]{Fig.~\ref{#1}}

\newcounter{topiccounter}
\renewcommand{\v}[1]{{\vec {#1}}}
\newcommand{\unit}[1]{\hat {\mathbf{#1}}} 

\newcommand{\e}{\varepsilon}
\newcommand{\aver}[1]{\left\langle #1 \right\rangle}
\begin{document}

\preprint{RBRC-856}

\title{Photon decay in strong magnetic field in heavy-ion collisions}

\author{Kirill Tuchin$\,^{a,b}$\\}

\affiliation{
$^a\,$Department of Physics and Astronomy, Iowa State University, Ames, IA 50011\\
$^b\,$RIKEN BNL Research Center, Upton, NY 11973-5000\\}

\date{\today}


\begin{abstract}

We calculate the photon pair production rate in strong magnetic field created in off-central heavy-ion collisions. Photon decay leads to depletion of the photon yield by a few percent at RHIC and by as much as 20\% at the LHC. It also generates a substantial azimuthal asymmetry (``elliptic flow") of the final photon distribution. We estimate $v_2\approx 2$\% at RHIC and $v_2\approx 14$\% at LHC.   Photon decay measurements is an important tool for studying the magnetic fields in early stages of heavy-ion collisions. 

\end{abstract}

\maketitle


Ultra-relativistic heavy ions colliding at finite impact parameter possibly create a super-critical magnetic field $B$. According to the estimates in \cite{Kharzeev:2007jp,Skokov:2009qp}, the strength of this field at $\sqrt s= 200$~GeV is about $eB\approx m_\pi^2/\hbar$, while the critical field is $eB_c=m_e^2/\hbar$. Thus,  magnetic field created in heavy-ion collisions is by many orders of magnitude  stronger than any field that has been created using the state-of-the-art lasers (see e.g.\ \cite{Marklund:2006my}). Possible existence of such  fields opens a new avenue for studying the high intensity regime of QED. 

Various QED processes in external magnetic field strongly depend on the time-dependence of that field. Recently we argued \cite{Tuchin:2010vs}  that the magnetic field is approximately stationary during the life-time of the quark-gluon plasma (QGP) that is formed shortly after the collision. Indeed, phenomenological models describing evolution of QGP indicate that the thermalized medium is formed almost immediately after the collision (after $\sim 0.5$~fm \cite{Kolb:2002ve,Kolb:2003dz}) when the magnetic field is near the maximum of its strength. As the heavy-ion remnants recede from the collision point, the magnetic field tends to rapidly decrease with time. This induces circular currents in the QGP that, by the Faraday law, produce an induced magnetic field  in the direction of the external field. Thus, the relaxation process of the external field slows down. The characteristic relaxation time is \cite{Tuchin:2010vs}
\beql{t-relax}
\tau = \frac{R^2\sigma}{4}\,,
\eeq
where $R$ is the QGP size and $\sigma$ is its electric conductivity.   
Lattice calculations show that the electric conductivity is high even at temperatures close to $T_c$ \cite{Gupta:2003zh}.  In \cite{Tuchin:2010vs} we used the lattice data of Ref.~\cite{Gupta:2003zh} to estimate the relaxation time as
 $\tau \approx 2.2\,\text{fm}\,(T/T_c)^2$. This number is even larger if the effect of magnetic field on the electric conductivity is taken into account \cite{Buividovich:2010tn}. It implies that external magnetic field is a  slowly varying function of  time during the entire QGP life-time.
 Of course, once the plasma cools down to the critical temperature and undergoes the phase transition to the hadronic gas, the conductivity becomes very small and the magnetic field cannot be sustained anymore.

In \cite{Tuchin:2010vs} we discussed the properties of  the synchrotron radiation of gluons by fast quarks and argued that it has a significant phenomenological implications. Indeed, the corresponding energy loss in magnetic field is comparable to that sustained by the fast quark in hot nuclear medium. The azimuthally asymmetric form of the energy loss contributes to the `elliptic flow' phenomenon observed at RHIC. In this letter we consider 
a cross-channel process -- pair-production by photon in external magnetic field. 
Specifically, we are interested to determine photon decay rate $w$ in the process $\gamma B\to f\bar f B$, where  $f$ stands for a charged fermion, as a function of photon's transverse momentum $k_T$, rapidity $\eta$ and azimuthal angle $\varphi$. Origin of these photons in heavy-ion collisions will not be of interest in this paper.

Characteristic frequency of a fermion  of species $a$ of mass  $m_a$ and charge $z_ae$ ($e$ is the absolute value of electron charge) moving in external magnetic field $B$ (in a plane perpendicular to the field direction) is 
\beql{char}
\hbar\omega_B= \frac{z_aeB}{\e}\,,
\eeq
where $\e$ is the fermion energy. Here -- in the spirit of the adiabatic approximation -- $B$ is a slow function of time. Calculation of the photon decay probability significantly simplifies if motion of electron is quasi-classical, i.e.\ quantization of fermion motion in the magnetic field can be neglected. This condition is fulfilled if $\hbar \omega_B\ll \e$. This implies that
\beql{est1}
\e\gg  \sqrt{zeB}\,.
\eeq
For RHIC it is equivalent to $\e\gg m_\pi$, for LHC $\e\gg 4m_\pi$.

Photon decay rate was calculated in \cite{Nikishov:1964zza} and, using the quasi-classical method, in \cite{Baier:1964}. It reads
\beql{main}
w=-\sum_a\frac{\alpha_\mathrm{em}\,z_a^3 \,e B }{m_a \varkappa_a}\int^\infty_{(4/ \varkappa_a)^{2/3}}\frac{2(x^{3/2}+1/ \varkappa_a)\,\text{Ai}'(x)}{x^{11/4}(x^{3/2}-4/ \varkappa_a)^{3/2}}\,,
\eeq
where summation is over fermion species and the invariant parameter $\varkappa$ is defined as 
\beq\label{chi}
 \varkappa_a^2 =-\frac{\alpha_\mathrm{em}z_a^2\hbar^3}{m_a^6}\,(F_{\mu\nu}k^\nu)^2 = \frac{\alpha_\mathrm{em}z_a^2\hbar^3}{m_a^6}(\v k\times \v B)^2\,,
\eeq
with the initial photon 4-momentum  $k^\mu=(\hbar \omega,\v k)$.  In heavy-ion collisions the vector of magnetic field $\vec B$ is orthogonal to the ``reaction plane", which is  spanned by the impact parameter $\vec b$ and  the collision axis $\hat z$. We define the  polar angle $\theta$ with respect to the $z$-axis and azimuthal angle $\varphi$ with respect to the reaction plane. 
In this notation, $\v B= B\,\unit y$ and $\v k = k_z\unit z+k_\bot (\unit x\, \cos\varphi+\unit y\sin\varphi)$, where $ k_\bot = |\v k|\sin\theta=\hbar \omega  \sin\theta$. Thus, $(\v B\times\v k)^2= B^2(k_z^2+k_\bot^2\cos^2\varphi)$. Introducing  rapidity $\eta$ as usual $\hbar \omega = k_\bot \cosh\eta$ and $k_z= k_\bot \sinh\eta$
we can write
\beql{chi-hi}
 \varkappa_a= \frac{\hbar(z_a e B)}{m_a^3}\,k_\bot \sqrt{\sinh^2\eta+\cos^2\varphi}\,.
\eeq

\begin{figure}[ht]
\begin{tabular}{cc}
      \includegraphics[height=5cm]{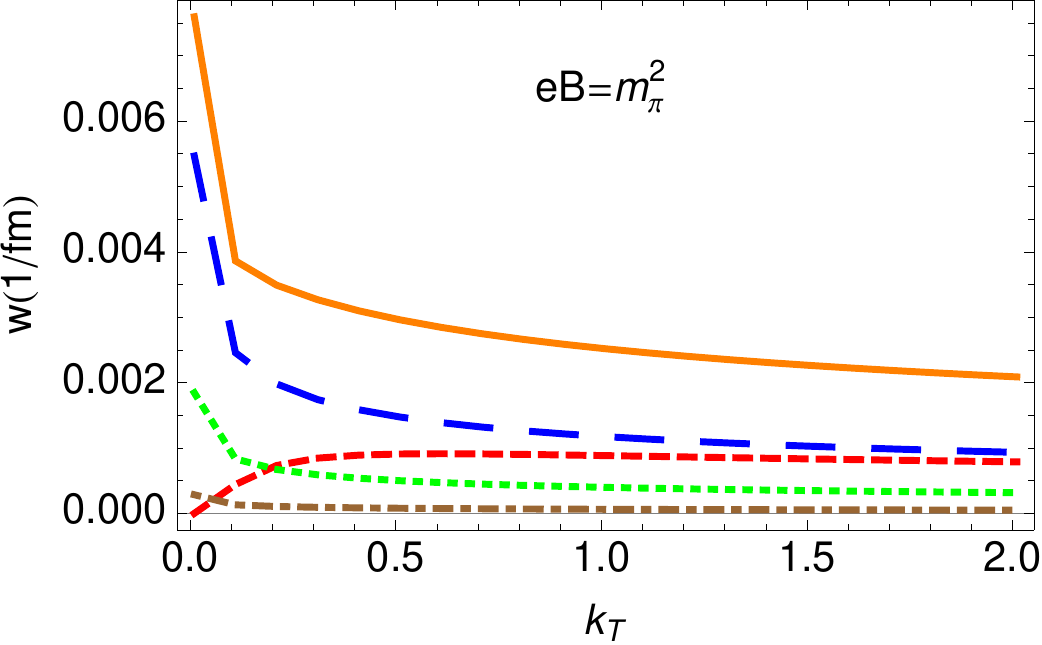} &
      \includegraphics[height=5cm]{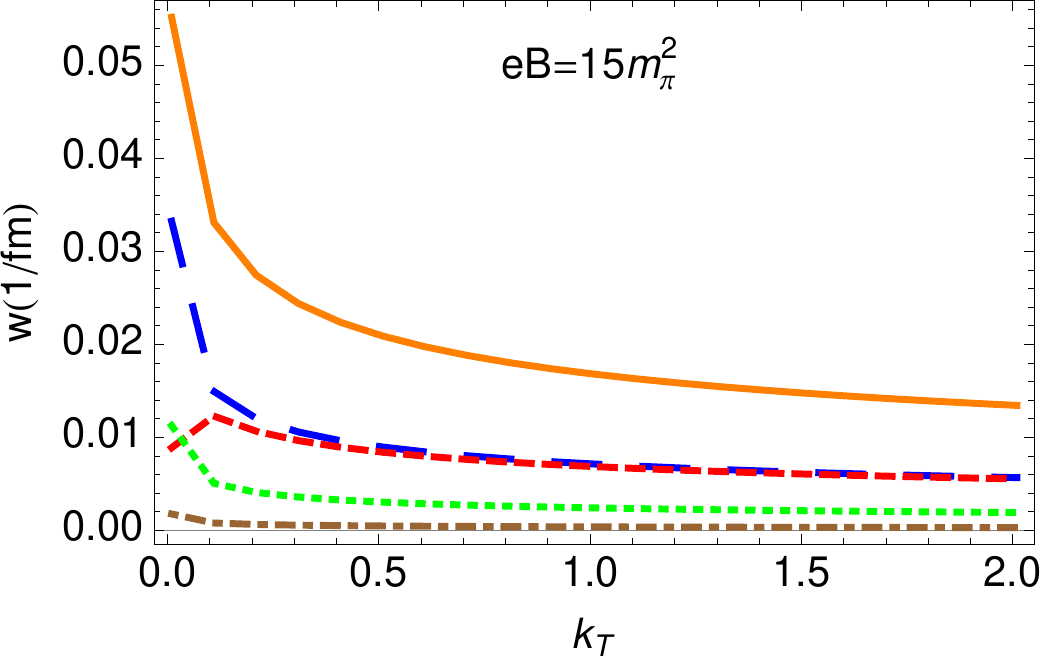}\\
      (a) & (b)
   \end{tabular}   
  \caption{Decay rate of photons moving in reaction plane in magnetic field as a function of transverse momentum $k_T$: (a) at RHIC, (b) at LHC. Broken lines from bottom to top give contributions of $\gamma\to d\bar d$, $\gamma\to u\bar u$, $\gamma\to \mu^+\mu^-$ and  $\gamma\to e^+e^-$ channels. Upper solid line is the total rate.   }
\label{rate-kt}
\end{figure}

In \fig{rate-kt} we plotted the  photon decay rate  \eq{main} for RHIC and LHC.
The survival probability of photons in magnetic field is $P= 1-w\Delta t $, where $\Delta t$ is the time spent by a photon in plasma.   We can see that for $\Delta t = 10$~fm the photon survives with probability $P_\text{RHIC}\approx 97$\% at RHIC, while at LHC $P_\text{LHC}\approx 80$\%. Such strong depletion can certainly be observed in heavy-ion collisions at LHC. 

\begin{figure}[ht]
      \includegraphics[height=8cm]{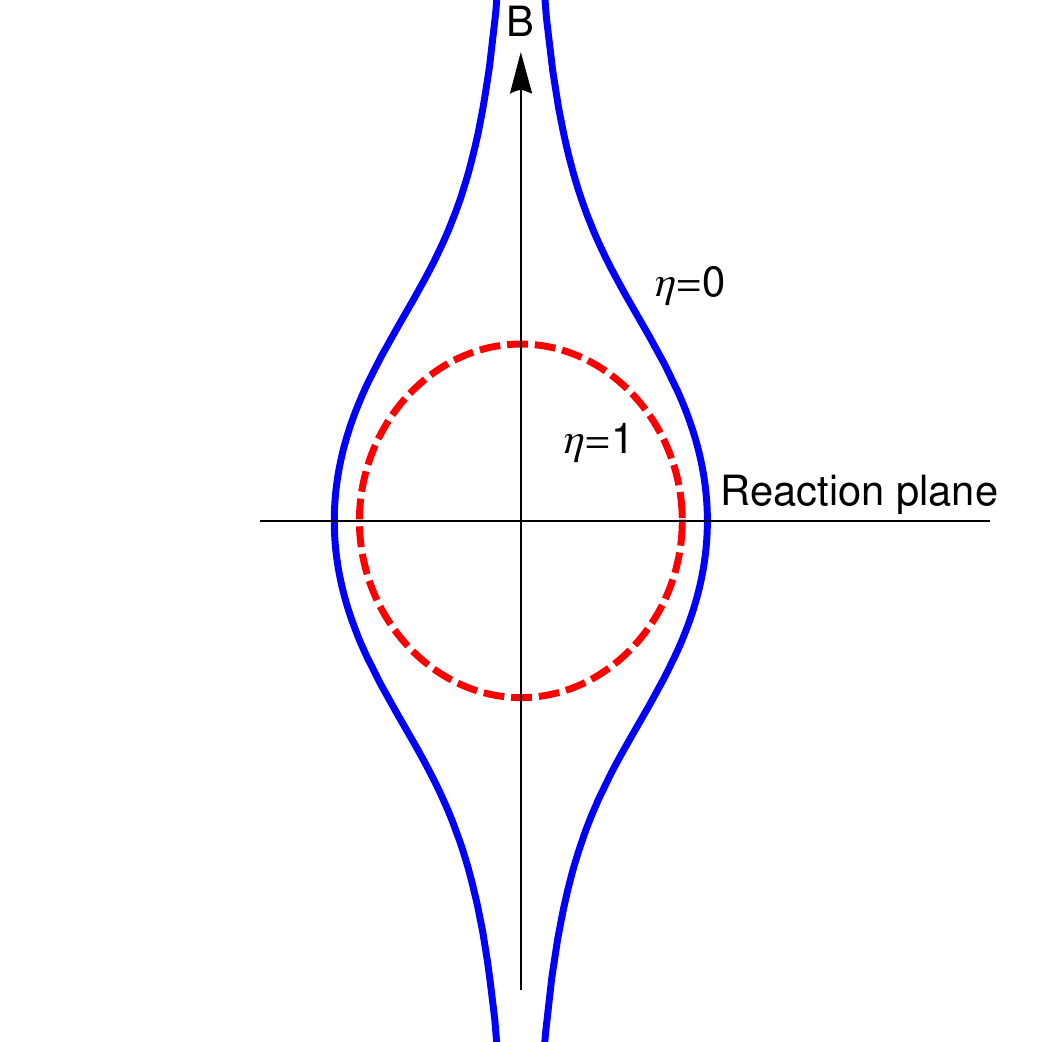} 
  \caption{Azimuthal distribution of the decay rate of photons at different rapidities at LHC. Only contribution of the $\gamma\to e^+e^-$ channel is shown.   }
\label{rate-azimuth}
\end{figure}

Azimuthal distribution of the decay rate of photons at LHC is azimuthally asymmetric as can be seen in \fig{rate-azimuth}. The strongest suppression is in the $B$ field direction, i.e.\ in the direction orthogonal to the reaction plane. At $\eta\gtrsim 1$ the $\varphi$ dependence of $\varkappa_a$ is very weak which is reflected in nearly symmetric azimuthal shape of the  dashed line in \fig{rate-azimuth}.

To quantify the azimuthal asymmetry it is customary to expand
the decay rate in Fourier series with respect to the azimuthal angle. Noting that $w$ is an even function of $\varphi$ we have
\beql{four}
w(\varphi)= \frac{1}{2}w_0+\sum_{n=1}^\infty w_n\,\cos(n\varphi)\,,\quad w_n= \frac{1}{\pi}\int_{-\pi}^{\pi}w(\varphi)\,\cos(n\varphi)\,d\varphi\,.
\eeq
In strong fields  $\varkappa_a\gg 1$. For example, for $\gamma\to \mu^+\mu^-$ at RHIC at $\varphi=\eta=0$ and $k_T=1$~GeV we get $\varkappa_\mu = 19$. Therefore, we can expand the rate  \eq{main}  at large $\varkappa_a$ as \cite{Nikishov:1964zza}
\beql{appr}
w\approx \frac{3^{1/6}\, 5\, \Gamma^2\left(\frac{2}{3}\right)}{2^{4/3}\,7\, \pi^{1/2}\, \Gamma\left(\frac{7}{6}\right)}\sum_a\frac{\alpha_\mathrm{em}eB z_a^3}{m_a\varkappa_a^{1/3}}\equiv \frac{A}{(\sinh^2\eta +\cos^2\varphi)^{1/6}}\,,\quad \varkappa_a \gg 1\,.
\eeq
At $\eta =0$ the Fourier coefficients $w_n$ can be calculated analytically using formula 3.631.9 of \cite{GR}
\beql{fc}
w_{2k}=\frac{3\,2^{1/3}\,A}{B\left( \frac{5}{6}+k,\frac{5}{6}-k\right)}\,,\quad   w_{2k+1}=0\,,\quad k=0,1,2,\ldots\,,
\eeq
where $B$ is the Euler's Beta-function and $A$ is defined in \eq{appr}. 
Substituting these expressions into \eq{four} we find
\beql{f22}
w=\frac{1}{2}w_0\left[ 1-\sum_{k=1}^\infty \frac{\sqrt{\pi}\Gamma\left(-\frac{1}{6}\right)}{2^{2/3}B\left( \frac{5}{6}+k,\frac{5}{6}-k\right)}\cos(2k\varphi)\right]
\eeq
The first few terms in this expansion read
\beql{f1}
w= \frac{1}{2}w_0\left( 1-\frac{2}{5}\cos(2\phi)+\frac{14}{55}\cos(4\phi)-\ldots\right)\,,
\eeq

What is measured experimentally is not the decay rate, but rather the photon spectrum. 
This spectrum is modified by the survival probability $P$ which is obviously azimuthally asymmetric. To quantify this asymmetry we write using \eq{four}
\beql{pphi}
P= \bar P \left( 1+ \sum_{k=1}^\infty v_{2k}\cos(2\varphi k)\right)\,,\quad v_{2k}= -\frac{1-\bar P}{\bar P}\,\frac{2\,w_{2k}}{w_0}\,,
\eeq
where $\bar P = \aver{1-w \Delta t}_\varphi=1-w_0\Delta t$ is the survival probability averaged over the azimuthal angle. Since $w_0\Delta t \ll 1$, as can be seen in \fig{rate-kt}, we can estimate using \eq{appr} and \eq{fc}
\beql{vs}
v_{2k}\approx  - \frac{2w_{2k}}{w_0}\,w_0 \Delta t =- \frac{2w_{2k}}{w_0}\Delta t\,\frac{5\, 6^{2/3}\Gamma\left(\frac{2}{3}\right)}{7\pi}\,\sum_a\frac{\alpha_\mathrm{em} (eB)^{2/3}z_a^{8/3}}{(k_T)^{1/3}}\,.
\eeq
In particular, the ``elliptic flow" coefficient is
\beql{v2}
v_2= \Delta t\,\frac{2\, 6^{2/3}\Gamma\left(\frac{2}{3}\right)}{7\pi}\,\sum_a\frac{\alpha_\mathrm{em} (eB)^{2/3}z_a^{8/3}}{(k_T)^{1/3}}\,.
\eeq
For example, at $k_T=1$~GeV and $\Delta t \sim 10$~fm/c   one expects $v_2\simeq   2$\% at RHIC and  $v_2  \simeq 14$\% at LHC only due to the presence of the magnetic field.  
We see that decay of photons in external magnetic field significantly contributes to the photon asymmetry in heavy-ion collisions  along with other possible effects \cite{Alam:1993jt,Turbide:2005bz,Chatterjee:2005de,Layek:2006um,Kopeliovich:2007sd,Kopeliovich:2007fv}.

In summary,  we calculated photon pair-production rate in external magnetic field created in off-central heavy-ion collisions. Photon decay leads to depletion of the photon yield by a few percent at RHIC and by as much as 20\% at the LHC. The decay rate depends on the rapidity and azimuthal angle. At mid-rapidity the azimuthal asymmetry of the decay rate translates into asymmetric photon yield and contributes to the ``elliptic flow". 
Let us also note that photons polarized parallel to the field are 3/2 times more likely to decay than those polarized transversely \cite{Nikishov:1964zza}. Therefore, polarization of the final photon spectrum perpendicular to the field is a signature of existence of the strong magnetic field. Finally, photon decay leads to enhancement of dilepton yield, which 
will be addressed in a separate publication.

\acknowledgments
I  am grateful to Dima Kharzeev for many informative discussions. 
 This work  was supported in part by the U.S. Department of Energy under Grant No.\ DE-FG02-87ER40371. I would like to
thank RIKEN, BNL, and the U.S. Department of Energy (Contract No.\ DE-AC02-98CH10886) for providing facilities essential
for the completion of this work.



\end{document}